\documentclass[aps,prl,reprint,showpacs,superscriptaddress]{revtex4-1}

\usepackage{graphicx,color,rotating,pifont}
\usepackage{amsmath,amssymb,bm}
\usepackage{ae}
\usepackage{pstricks}
\DeclareMathSymbol{\NS}{\mathord}{AMSb}{"4E}

\DeclareMathOperator{\sgn}{\mathrm{sgn}}

\newcommand{\ket}[1]{\ensuremath{\,|{#1}\rangle}}

\newcommand{\matrixe}[3]{\ensuremath{\langle{#1}|\,{#2}\,|{#3}\rangle}}

\newcommand{\dmatrixe}[2]{\matrixe{#1}{#2}{#1}}

\newcommand{\comm}[2]{\ensuremath{[{#1},{#2}]}}

\newcommand{\op}[1]{\ensuremath{#1}}
\newcommand{\adj}[1]{\ensuremath{{{#1}}^{\dag}}}

\newcommand{\totd}[2]{\ensuremath{ \frac{d {#1}} {d {#2}} }}

\newcommand{\nn}{\ensuremath{\bar{n}}}

\newcommand{\aO}{\ensuremath{\op{a}}}

\newcommand{\etaO}{\ensuremath{\op{\eta}}}

\newcommand{\aaO}{\ensuremath{\adj{\op{a}}}}

\newcommand{\AO}{\ensuremath{\op{A}}}

\newcommand{\HO}{\ensuremath{\op{H}}}

\newcommand{\PO}{\ensuremath{\op{P}}}

\newcommand{\lambdaSRG}{\ensuremath{\lambda_{\text{SRG}}}}

\newcommand{\EMax}{\ensuremath{E_{3\text{max}}}}

\newcommand{\hw}{\ensuremath{\hbar\omega}}

\newcommand{\nuc}[2]{\ensuremath{^{#2}\mathrm{#1}}}

\newcommand{\fm}{\ensuremath{\,\text{fm}}}
\newcommand{\fmi}{\ensuremath{\fm^{-1}}}

\newcommand{\keV}{\ensuremath{\,\text{keV}}}
\newcommand{\MeV}{\ensuremath{\,\text{MeV}}}

\newcommand{\nord}[1]{\ensuremath{:\!#1:}}

\newcommand{\NNNLO}{N$^3$LO}

\definecolor{FGViolet}{rgb}{0.61,0.32,0.61}
\definecolor{FGPurple}{rgb}{0.31,0.18,0.31}
\definecolor{FGDarkBlue}{rgb}{0,0,0.6}
\definecolor{FGBlue}{rgb}{0,0,0.8}
\definecolor{FGLightBlue}{rgb}{0.2, 0.6, 0.8}
\definecolor{FGGreen}{rgb}{0.2,0.7,0.2}
\definecolor{FGLightGreen}{rgb}{0.4,1,0.4}
\definecolor{FGYellow}{rgb}{1,0.95,0}
\definecolor{FGOrange}{rgb}{0.95,0.5,0.1}
\definecolor{FGRed}{rgb}{0.8,0,0}
\definecolor{FGWhite}{rgb}{1,1,1}
\definecolor{FGLightGray}{rgb}{0.8,0.8,0.8}
\definecolor{FGGray}{rgb}{0.5,0.5,0.5}
\definecolor{FGDarkGray}{rgb}{0.3,0.3,0.3}
\definecolor{FGBlack}{rgb}{0,0,0}

\begin{document}
\title{Ab Initio Multi-Reference In-Medium Similarity Renormalization Group Calculations 
of Even Calcium and Nickel Isotopes}

\author{H.\ Hergert}
\email{Corresponding author. E-mail: hergert.3@osu.edu}
\affiliation{The Ohio State University, Columbus, OH 43210, USA}

\author{S.\ K.\ Bogner}
\affiliation{National Superconducting Cyclotron Laboratory
and Department of Physics and Astronomy, Michigan State University,
East Lansing, MI 48844, USA}

\author{T.\ D.\ Morris}
\affiliation{National Superconducting Cyclotron Laboratory
and Department of Physics and Astronomy, Michigan State University,
East Lansing, MI 48844, USA}

\author{S.\ Binder}
\affiliation{Institut f\"ur Kernphysik, Technische Universit\"at Darmstadt,
D-64289 Darmstadt, Germany}

\author{A.\ Calci}
\affiliation{Institut f\"ur Kernphysik, Technische Universit\"at Darmstadt,
D-64289 Darmstadt, Germany}

\author{J.\ Langhammer}
\affiliation{Institut f\"ur Kernphysik, Technische Universit\"at Darmstadt,
D-64289 Darmstadt, Germany}

\author{R.\ Roth}
\affiliation{Institut f\"ur Kernphysik, Technische Universit\"at Darmstadt,
D-64289 Darmstadt, Germany}

\date{\today}

\begin{abstract}
We use the newly developed Multi-Reference In-Medium Similarity Renormalization Group to study all even isotopes of the calcium and nickel isotopic chains, based on two- plus three-nucleon interactions derived from chiral effective field theory. We present results for ground-state and two-neutron separation energies and quantify their theoretical uncertainties. At shell closures, we find excellent agreement with Coupled Cluster results obtained with the same Hamiltonians. Our results confirm the importance of chiral 3N interactions to obtain a correct reproduction of experimental energy trends, and their subtle impact in neutron-rich Ca and Ni isotopes. At the same time, we uncover and discuss deficiencies of the input Hamiltonians which need to be addressed by the next generation of chiral interactions. 
\end{abstract}


\maketitle

\clearpage

\paragraph{Introduction.}
As experimental capabilities for the production of rare isotopes grow, so does the need for a reliable description and prediction of their properties from nuclear many-body theory, including quantified uncertainties. Systematically improvable \emph{ab initio} methods like Coupled Cluster (CC) \cite{Hagen:2014ve,Binder:2014fk,Henderson:2014ly}, Self-Consistent Green's Functions \cite{Dickhoff:2004fk,Cipollone:2013uq,Soma:2011vn,Soma:2013ys,Soma:2014eu}, and the In-Medium Similarity Renormalization Group (IM-SRG) \cite{Tsukiyama:2011uq,Hergert:2013mi,Hergert:2013ij} routinely access medium-mass closed-shell nuclei and even heavy systems beyond $A=100$ due to their modest computational scaling. At the same time, great effort has been invested to quantify the theoretical uncertainties of these methods \cite{Binder:2013zr,Hergert:2013mi,Roth:2014fk,Binder:2014fk}.

Nuclear interactions from chiral effective field theory are the input of choice for \emph{ab initio} many-body theory, because they provide formally consistent two-, three-, and up to $A$-nucleon forces and operators (see, e.g., \cite{Epelbaum:2009ve,Hammer:2013nx}). Current chiral Hamiltonians have been employed with great success, but there are open issues regarding the power counting, the determination of the low-energy constants (LECs) etc., motivating a concerted effort to construct next-generation chiral interactions for the nuclear structure and reactions community \cite{Lenpic}.

\emph{Ab initio} studies of medium-mass and heavy nuclei, particularly away from closed shells, allow us to confront chiral Hamiltonians with a wealth of experimental data from existing and forthcoming rare-isotope facilities. Such nuclei are sensitive to features of the Hamiltonian which are not probed in few-body systems, and exotic nuclei, in particular, are an excellent laboratory to study the interplay of the two-nucleon (NN) and three-nucleon (3N) interactions, as well as continuum effects. For instance, chiral 3N forces have been crucial for a proper description of the neutron drip lines in the region around oxygen \cite{Otsuka:2010cr,Hagen:2012oq,Hagen:2012nx,Holt:2013hc,Hergert:2013ij,Cipollone:2013uq,Soma:2014eu,Bogner:2014tg,Jansen:2014qf}. 

In this work, we study the even calcium and nickel isotopes with the Multi-Reference IM-SRG (MR-IM-SRG) for open-shell nuclei, using chiral NN+3N interactions as input. Such a study is timely, because neutron-rich calcium isotopes have been the focus of ongoing experimental campaigns \cite{Lapierre:2012fu,Gallant:2012kx,Steppenbeck:2013dq,Wienholtz:2013bh,Gade:2014fk,Riley:2014dz}, and investigations of proton-rich Ca isotopes are planned for the near future. Likewise, there is continued interest in neutron-rich Ni isotopes \cite{Mazzocchi:2005nx,Hakala:2008ij,Aoi:2010cr,Broda:2012zr,Chiara:2012ys,Zhu:2012tg,Recchia:2013kx,Tsunoda:2014ve}. 

\paragraph{Multi-Reference In-Medium SRG.}
The basic formalism of the (MR)-IM-SRG is presented in Refs.~\cite{Tsukiyama:2011uq,Hergert:2013mi,Hergert:2013ij}. The Hamiltonian is normal-ordered with respect to an arbitrary reference state $\ket{\Phi}$ via the generalized normal-ordering developed by Kutzelnigg and Mukherjee \cite{Kutzelnigg:1997fk,Kong:2010kx}, and plugged into the operator flow equation
\begin{equation}\label{eq:flow}
  \totd{}{s}\HO(s) = \comm{\etaO(s)}{\HO(s)}\,.
\end{equation}
With a suitable choice of generator $\etaO(s)$, Eq.~\eqref{eq:flow} implements a continuous unitary transformation that decouples the ground state of the Hamiltonian $\HO(s)$ from excitations as we evolve $s\to\infty$, solving the many-body problem \cite{Tsukiyama:2011uq, Hergert:2013mi,Hergert:2013ij}. We close the system of flow equations by truncating $\eta(s)$ and $\HO(s)$ at the two-body level for all $s$, obtaining the scheme we refer to as MR-IM-SRG(2). 

For an arbitrary reference state $\ket{\Phi}$, the flow equations do not only depend on the one-body density matrix 
$\lambda^{1}_{2} \equiv \dmatrixe{\Phi}{\aaO_1\aO_2}$~\cite{Ring:1980bb}, but also on irreducible two-,\ldots,$A$-body density matrices, which encode information on the correlations in the state \cite{Kutzelnigg:1997fk,Kong:2010kx,Mukherjee:2001uq}. They are defined recursively by subtracting reducible contributions from the full $n$-body density matrices, e.g.,
\begin{align}
  \lambda^{12}_{34} \equiv \dmatrixe{\Phi}{\!\AO^{12}_{34}} - \lambda^{1}_{2}\lambda^{3}_{4} 
                       + \lambda^{1}_{3}\lambda^{2}_{4}\,,
\end{align}
where $\AO^{1\ldots k}_{l\ldots N}\equiv\aaO_1\ldots\aaO_k\aO_N\ldots\aO_l$ is a compact notation for strings of creation and annihilation operators. 

The MR-IM-SRG reference state is a Hartree-Fock-Bogoliubov (HFB) quasiparticle vacuum which has been projected on the proton and neutron number of the target nucleus, $\ket{\Phi}=\PO_N\PO_Z\ket{\text{HFB}}$~\cite{Hergert:2013ij}. For such states, it is sufficient to truncate terms in the flow equations that are non-linear in $\lambda^{12}_{34}$ or contain irreducible three-body density matrices: Energy changes due to this truncation are negligible compared to other sources of uncertainty discussed in the following.

In the present work, we use the notation of Ref.~\cite{Hergert:2013ij} but a different ansatz for the generator $\etaO(s)$. Its matrix elements are defined as ($\lambda^i_k=n_i\delta^i_k,\, \bar{n}_i=1-n_i$)
\begin{align}
  \eta^{1}_{2} &\equiv \sgn(\Delta^{1}_{2})\nn_1 n_2 f^1_2 - \left[1 \leftrightarrow 2 \right]\,,\label{eq:eta1b}\\
  \eta^{12}_{34} &\equiv \sgn(\Delta^{12}_{34})\nn_1\nn_2 n_3n_4 \Gamma^{12}_{34} - \left[(12) \leftrightarrow (34) \right]\,,\label{eq:eta2b}
\end{align}
where $f$ and $\Gamma$ are the one- and two-body parts of the normal-ordered Hamiltonian $\HO$. Indicating normal ordering by colons, the expressions
\begin{align}
  \Delta^{1}_{2} &\equiv \matrixe{\Phi}{\nord{A^{2}_{1}}\HO\nord{A^{1}_{2}}}{\Phi} - \matrixe{\Phi}{\HO}{\Phi}\,, \label{eq:def_Delta1b}\\
  \Delta^{12}_{34} &\equiv \matrixe{\Phi}{\nord{A^{34}_{12}}\HO\nord{A^{12}_{34}}}{\Phi} - \matrixe{\Phi}{\HO}{\Phi}\,\label{eq:def_Delta2b}
\end{align}
are evaluated using the generalized Wick theorem \cite{Kutzelnigg:1997fk,Kong:2010kx}, truncating irreducible densities as in the flow equations. This generator suppresses the off-diagonal one- and two-body matrix elements that couple to the reference state $\ket{\Phi}$ to excitations like $e^{-|\Delta|s}$ as $s\to\infty$, where $\Delta$ is the corresponding energy difference \eqref{eq:def_Delta1b} or \eqref{eq:def_Delta2b}. We refer to $\eta$ as the imaginary-time generator due to its relation to imaginary-time projection operators as used, e.g., in Green's Function Monte Carlo \cite{Kalos:1962jl,Carlson:1987ys}. It is of similar efficiency as the White-type generator used in Ref.~\cite{Hergert:2013ij} due to its low construction cost and the moderate stiffness of the flow equations it generates, but does not suffer from instabilities due to small energy denominators.

\paragraph{Hamiltonians and Implementation.}
In this work, we use the chiral \NNNLO \ NN interaction by Entem and Machleidt, with non-local cutoff $\Lambda_\text{NN}=500\,\MeV/c$ \cite{Entem:2002sd,Machleidt:2011bh}. Where indicated, it will be accompanied by a local N$^2$LO 3N interaction with initial cutoffs $\Lambda_\text{3N}=350\,$ and $400\,\MeV/c$ \cite{Roth:2012qf,Hergert:2013ij,Roth:2014fk,Binder:2013zr}. The reduced values of $\Lambda_\text{3N}$ avoid strong induced 4N interactions if this Hamiltonian is softened via free-space SRG evolution \cite{Roth:2012qf,Roth:2014fk}. While $\Lambda_\text{NN}$ and $\Lambda_\text{3N}$ are nominally inconsistent, we note that the NN and 3N interactions are regularized in different schemes, so their values should not necessarily be the same \cite{Hagen:2014fk}. This issue will be revisited with a new generation of consistently regularized chiral Hamiltonians in the future \cite{Lenpic}.

The Hamiltonians are softened by a free-space SRG evolution at the three-body level to $\lambdaSRG=1.88,\ldots,2.24\,\fm^{-1}$ \cite{Jurgenson:2009bs,Roth:2011kx,Roth:2014fk}. Hamiltonians that only contain SRG-induced 3N forces are referred to as NN+3N-induced, those also containing an initial 3N interaction as NN+3N-full. 

Working with harmonic oscillator (HO) single-particle states, we truncate the 3N matrix elements in the total energy quantum number according to $e_1+e_2+e_3\leq\EMax$, due to memory requirements \cite{Roth:2012qf,Binder:2013zr,Hergert:2013mi,Hergert:2013ij,Binder:2014fk}. Uncertainties caused by this truncation are investigated below. 
As discussed in Refs.~\cite{Binder:2014fk,Roth:2014fk}, the free-space SRG evolution of the input Hamiltonian must be performed in a sufficiently large model space for $pf$-shell and heavier nuclei. 
Therefore, we rely on the model space $\mathcal{B}$ from Ref.~\cite{Binder:2014fk} for the evolution of the 3N force. The oscillator parameter of the Jacobi HO basis is chosen to be $\hw=36\,\MeV$. Matrix elements for smaller $\hw$ are obtained by frequency conversion \cite{Roth:2014fk}.

Our results are converged with respect to the size of the single-particle basis: At the $\hbar\omega$ value of the energy minimum, the change in the ground-state energy is 0.1\% when we increase the basis from 13 to 15 major shells. 

To obtain reference states for the MR-IM-SRG(2), we solve HFB equations in 15 major HO shells, and project the resulting state on proton and neutron numbers \cite{Hergert:2009zn,Hergert:2013mi,Hergert:2013ij}. The intrinsic NN+3N Hamiltonian is normal-ordered with respect to the reference state, and the residual 3N interaction term is discarded. This normal-ordered two-body approximation (NO2B) is found to overestimate binding energies by less than 1\% in the calcium and nickel isotopes \cite{Roth:2012qf,Binder:2013zr,Hergert:2013mi,Binder:2014fk}.

\paragraph{Calcium isotopes.}

\begin{figure}[t]
  \includegraphics[width=0.95\columnwidth]{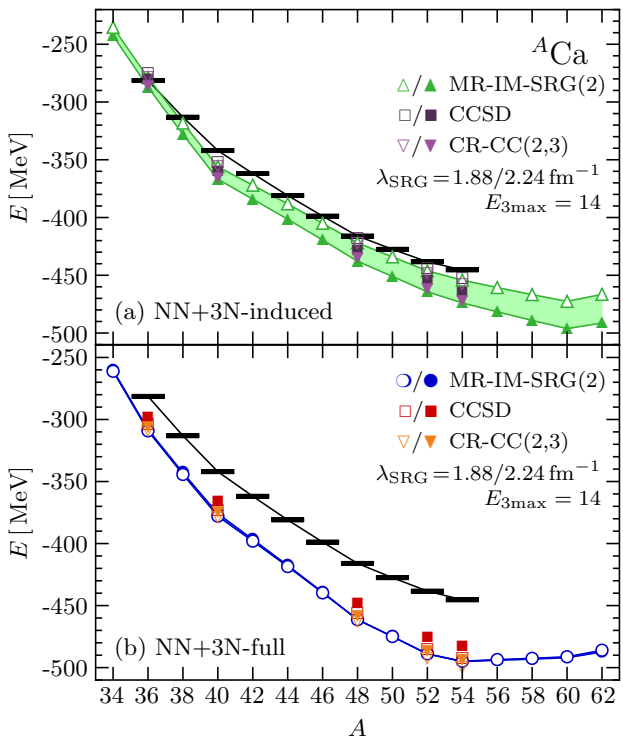}
  \vspace{-5pt}
\caption{\label{fig:Ca_Eint}(Color online) Ground-state energies of the Ca isotopes for the NN+3N-induced (a) and NN+3N full (b) Hamiltonians, with $\lambdaSRG=1.88\fmi$ (open symbols) to $2.24\fmi$ (solid symbols). The bands for the MR-IM-SRG(2) results indicate the variation of the results with the resolution scale $\lambdaSRG$. Experimental data (black bars) are taken from \cite{Wang:2012uq,Wienholtz:2013bh}.
}
\end{figure}

In Fig.~\ref{fig:Ca_Eint}, we show MR-IM-SRG(2) ground-state energies for $\nuc{Ca}{34-62}$, along with CC results including doubles (CCSD) \cite{Shavitt:2009} and triples excitations (CR-CC(2,3)) \cite{Piecuch:2005dp,Binder:2013fk} for closed-shell isotopes. Surveying the results, we note that MR-IM-SRG(2) and CR-CC(2,3) results are in very good agreement. It is a recurring theme that (MR-)IM-SRG(2) provides results comparable to CC approaches that include triples (3-particle, 3-hole) excitations at the computational cost of a doubles (2p2h) method \cite{Tsukiyama:2011uq,Hergert:2013mi,Hergert:2013ij}.

For the NN+3N-induced Hamiltonian shown in Fig.~\ref{fig:Ca_Eint}(a), we overbind the Ca isotopes for the considered values of $\lambdaSRG$. However, the ground-state energies vary significantly with the resolution scale $\lambdaSRG$ due to omitted induced beyond-3N forces. Other sources, such as the $\EMax$ truncation and NO2B approximation, can be ruled out because they are only weakly sensitive to $\lambdaSRG$ variations~\cite{Binder:2013zr,Hergert:2013mi,Hergert:2013ij,Binder:2014fk}. Furthermore, the $\lambdaSRG$ dependence of MR-IM-SRG(2) and CR-CC(2,3) is comparable despite their different many-body content, which implies that missing many-body effects cannot be its primary source, either. 

In Fig.~\ref{fig:Ca_Eint}(b), we show that the inclusion of an initial 3N force reduces the $\lambdaSRG$ dependence drastically. As discussed in Ref.~\cite{Binder:2014fk}, this is a result of cancellations between induced forces from the initial NN and 3N interactions. With this reduced dependence on $\lambdaSRG$ we find an overbinding that is robust under variations of $\lambdaSRG$ and slowly increasing from 8\% for \nuc{Ca}{36} to 12\% for \nuc{Ca}{54}. 

\begin{figure}[t]
  \includegraphics[width=0.95\columnwidth]{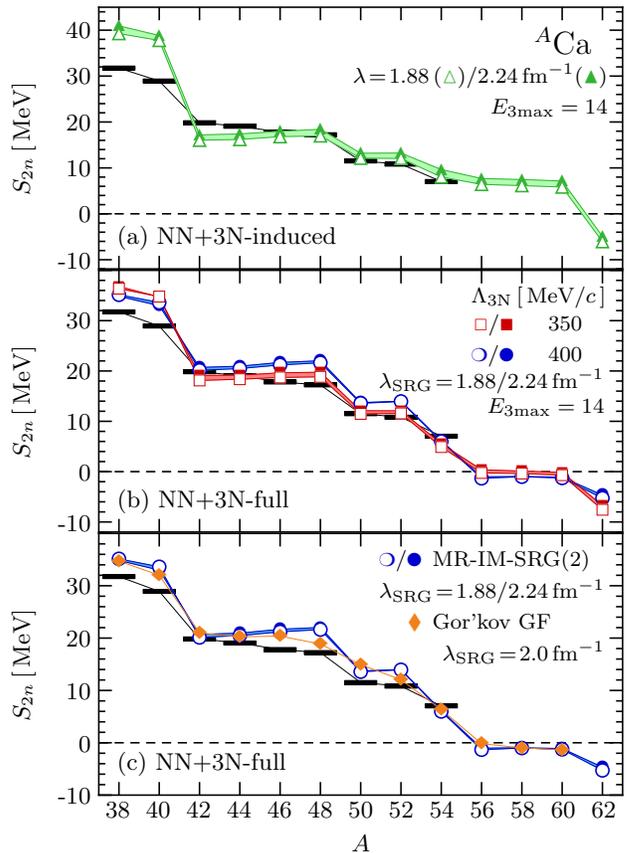}
  \vspace{-5pt}
\caption{\label{fig:Ca_S2N}(Color online) Two-neutron separation energies of the Ca isotopes for the NN+3N-induced (a) and NN+3N-full Hamiltonian with $\Lambda_\text{3N}=350,400\,\MeV/c$ (b), for a range $\lambdaSRG=1.88\fmi$ (open symbols) to $2.24\fmi$ (solid symbols). Panel (c) compares MR-IM-SRG(2) and second-order GGF \cite{Soma:2011vn,Soma:2013ys,Soma:2014eu} results with the same input Hamiltonian, but slightly different SRG evolution \cite{Soma:2014}. Experimental values (black bars) are taken from \cite{Wang:2012uq,Wienholtz:2013bh}.}
\end{figure}

We now consider the two-neutron separation energies $S_{2n}$ shown in Fig.~\ref{fig:Ca_S2N}. Such differential quantities filter out global energy shifts due to missing induced many-body forces, as well as many-body and basis truncations. For instance, the absolute variation of the $S_{2n}$ with $\lambdaSRG$ in the NN+3N-induced case is much weaker than the variation of the ground-state energies in Fig.~\ref{fig:Ca_Eint}(a).

The $S_{2n}$ for the NN+3N-induced Hamiltonian in Fig.~\ref{fig:Ca_S2N}(a) show a pronounced shell closure at $\nuc{Ca}{40}$, with $S_{2n}$ dropping by more than 20 MeV. The $\nuc{Ca}{48}$ shell closure is weak in comparison, albeit close to experimental data, and there are even weaker hints of shell closures in $\nuc{Ca}{52,54}$ (the reference states exhibit pairing in both cases). 
The $S_{2n}$ increase notably from $\nuc{Ca}{42}$ to $\nuc{Ca}{48}$, and weakly from $\nuc{Ca}{50}$ to $\nuc{Ca}{52}$. This is an indication that interaction components which are being accessed as neutrons are added to the $pf$-shell are too attractive, which is consistent with the observed overbinding. However, shell structure effects clearly also play a role, because the overbinding becomes less severe around $\nuc{Ca}{48}$ before increasing again with the neutron number $N$, while the $S_{2n}$ are always decreasing between shell closures beyond $\nuc{Ca}{52}$.

The NN+3N-induced Hamiltonian produces a distinct drip-line signal in Figs.~\ref{fig:Ca_Eint}(a) and \ref{fig:Ca_S2N}(a): $\nuc{Ca}{62}$ is consistently unbound by $5-6\,\MeV$ with respect to $\nuc{Ca}{60}$ for our range of $\lambdaSRG$. The change in $S_{2n}$ is much larger than the uncertainties due to many-body and basis truncations, or missing induced forces (see below). The inclusion of continuum effects in Ref.~\cite{Hagen:2012nx} reduced the energy of low-lying unbound states only by about $2\,\MeV$, which is insufficient to bind isotopes with $N>40$ with respect to $\nuc{Ca}{60}$. Without the inclusion of initial 3N forces, the drip line is therefore expected at $N=40$. 

In Fig.~\ref{fig:Ca_S2N}(b), we show $S_{2n}$ for NN+3N-full Hamiltonians with $\Lambda_\text{3N}=350,400\,\MeV/c$. The $N=20$ shell closure is weakened by the 3N forces, although the calculated $S_{2n}$ are still larger than experimental data. As before, we observe an increase of the separation energies for $\nuc{Ca}{42-48}$ and $\nuc{Ca}{50-52}$, but we note that the overbinding consistently increases with $N$ in this case (Fig.~\ref{fig:Ca_Eint}(b)). Interestingly, the $S_{2n}$ trends in these nuclei are flatter for $\Lambda_{\text{3N}}=350\MeV/c$ than for $400\MeV/c$, which suggests a change in the shell structure of these nuclei. Overall, the $S_{2n}$ are consistent under this variation of the 3N cutoff. In contrast to the NN+3N-induced case, both $\nuc{Ca}{52}$ and $\nuc{Ca}{54}$ exhibit magicity, in agreement with experimental and Shell Model results \cite{Gallant:2012kx,Holt:2012fk,Steppenbeck:2013dq,Wienholtz:2013bh,Holt:2014vn}. 

For large neutron numbers, the trends shown in Figs.~\ref{fig:Ca_Eint}(b) and \ref{fig:Ca_S2N}(b) are different from the NN+3N-induced case. $\nuc{Ca}{56-60}$ are unbound with respect to $\nuc{Ca}{54}$ by a mere $1-2\,\MeV$ (also see \cite{Hagen:2012nx}).
Consequently, these isotopes are sensitive to continuum effects and details of the interaction, which could lead to phenomena like neutron halos as proposed in \cite{Hagen:2013fk}. Figure~\ref{fig:Ca_S2N}(b) also shows that the flat plateau of the $S_{2n}$ for $\nuc{Ca}{56-60}$ in the vicinity of zero is remarkably robust under the variation of the cutoff of the initial 3N interaction from $400\,\MeV/c$ to $350\,\MeV/c$.

The Ca isotopes were also studied recently with the second-order Gor'kov Green's Function (GGF) method. The $S_{2n}$ published in Ref.~\cite{Soma:2014eu} were obtained with the same NN+3N-full Hamiltonian with $\Lambda_\text{3N}=400\MeV/c$, but a smaller 3N Jacobi HO model space was used for the SRG evolution than in our calculations. While the $S_{2n}$ systematics remain the same,    
we show updated GGF results \cite{[{}][{. For the GGF calculations, the NN+3N-full input Hamiltonian is SRG-evolved in a Jacobi HO model space with $E_\text{SRG}=44,42,42,40,38$ for $J=\tfrac{1}{2},\tfrac{3}{2},\tfrac{5}{2},\tfrac{7}{2},\tfrac{9}{2}$, respectively. From $J=\tfrac{11}{2}$ to $\tfrac{25}{2}$, $E_\text{SRG}$ is decreased in two-step plateaus from 36 to 30, and $E_\text{SRG}=16$ for $J>\tfrac{25}{2}$. Frequency conversion is not applied, i.e., the $\hw$ values of the Jacobi and single-particle HO bases are directly related through the coordinate transformation.}]Soma:2014} in Fig.~\ref{fig:Ca_S2N}(c) to allow a more quantitative comparison with our MR-IM-SRG(2) separation energies. The two methods agree well for mid-shell Ca isotopes, implying that the difference between second-order GGF and MR-IM-SRG(2) ground-state energies is primarily a global shift for these nuclei. Around shell closures, the broken particle-number symmetry in the GGF approach causes smoother trends due to pairing fluctuations (compare, e.g., HFB and number-projected HFB $S_{2n}$ in Ref.~\cite{Hergert:2008oe}). Overall, our results are consistent with the findings and conclusions of Ref.~\cite{Soma:2014eu}.

\begin{figure}
  \includegraphics[width=0.95\columnwidth]{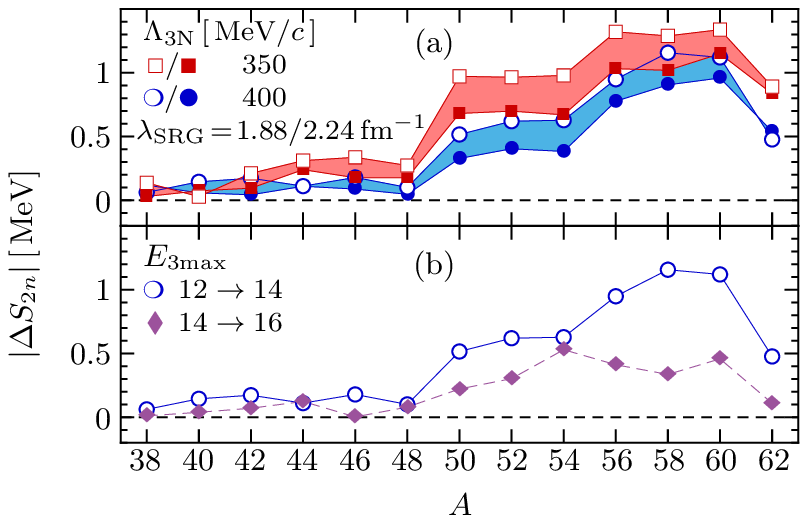}
\caption{\label{fig:Ca_DeltaS2N}(Color online) Uncertainty of Ca two-neutron separation energies: (a) variation as $\EMax=12\to14$ for different $\Lambda_\text{3N}$ and $\lambdaSRG=1.88$ to $2.24\fmi$. (b) Variation $\EMax=12\to14\to16$ for $\Lambda_\text{3N}=400\,\MeV/c$, $\lambdaSRG=1.88\fmi$.}
\end{figure}

Let us now discuss the uncertainties of the calcium two-neutron separation energies in more detail.
From the CC ground-state energies included in Fig.~\ref{fig:Ca_Eint} we can determine $S_{2n}$ in $\nuc{Ca}{54}$. We find a difference of only $150\,\keV$ between the CCSD and CR-CC(2,3) results, almost independent of $\Lambda_\text{3N}$ and $\lambdaSRG$.
 Due to the rapid convergence of the many-body expansion for soft Hamiltonians, we can interpret this as a measure for many-body effects that are not included in the MR-IM-SRG(2), by analogy with CC.

The uncertainties of the $S_{2n}$ due to the $\EMax$ truncation are explored further in Fig.~\ref{fig:Ca_DeltaS2N}. The contributions of the many-body and $\EMax$ truncations are of comparable size: Increasing $\EMax$ from our default 14 to 16, the $S_{2n}$ change by less than $100\,\keV$ for $\nuc{Ca}{38-50}$, and $200-500\,\keV$ for $\nuc{Ca}{52-60}$ (Fig.~\ref{fig:Ca_DeltaS2N}(b)). Comparing to the increase $\EMax=12\to14$, which causes variations as large as $1-1.5\,\MeV$ in the $S_{2n}$ of the mid- and upper $pf$-shell calcium isotopes, we see clear signs of convergence. 
Given the flat ground-state energy trend beyond $\nuc{Ca}{54}$ for the NN+3N-full Hamiltonians (cf.~Figs.~\ref{fig:Ca_Eint}~and~\ref{fig:Ca_S2N}), we conclude that our uncertainties are still too large to clearly identify the neutron dripline. A first step towards a more accurate calculation would be the exploration of $\EMax\geq16$~\cite{Binder:2014fk}.

\paragraph{Nickel isotopes.}

\begin{figure}[t]
  \includegraphics[width=0.95\columnwidth]{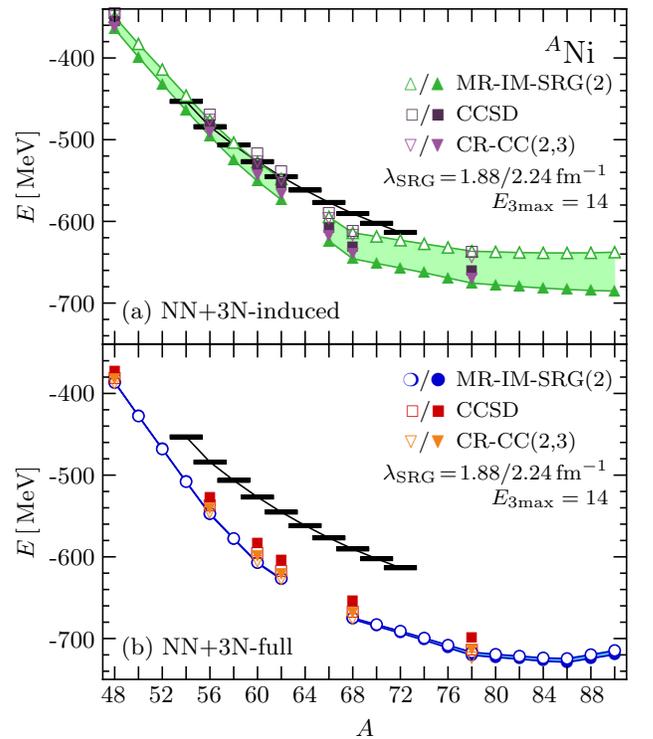}
\caption{\label{fig:Ni_Eint}(Color online) Ground-state energies of the Ni isotopes for the NN+3N-induced (a) and NN+3N full (b) Hamiltonians, for resolution scales $\lambdaSRG=1.88\fmi$ (open symbols) to $2.24\fmi$ (solid symbols). Experimental data (black bars) are taken from \cite{Wang:2012uq}.}
\end{figure}

We now focus on the nickel isotopes. Figure~\ref{fig:Ni_Eint} shows the ground-state energies for the NN+3N-induced and NN+3N-full Hamiltonians ($\Lambda_\text{3N}=400\,\MeV/c$). The basic features are very similar to the Ca case. MR-IM-SRG(2) and CC results are in very good agreement. The NN+3N-induced Hamiltonian (Fig.~\ref{fig:Ni_Eint}(a)) yields energies that are close to experimental binding energies for lighter Ni isotopes, but produces overbinding with growing neutron excess. The $\lambdaSRG$ dependence serves as an indicator for the size of missing 4N forces. With the inclusion of the initial 3N interaction (Fig.~\ref{fig:Ni_Eint}(b)), the overbinding is increased, while the $\lambdaSRG$ dependence is reduced due to cancellations between induced beyond-3N terms. Beyond $\nuc{Ni}{74}$, the ground-state energy curve becomes flat. In contrast to the Ca case, the NN+3N-induced also produces a very flat trend for these isotopes.

An apparent deformation instability emerges for $\nuc{Ni}{64}$ in the NN+3N-induced and $\nuc{Ni}{64,66}$ in the NN+3N-full cases. Because spherical symmetry is enforced in our calculations, we observe strong oscillations in the MR-IM-SRG(2) ground-state energy and the norm of the generator. Usually, the latter decreases monotonically until convergence.
Experimental spectra of these Ni isotopes show spherical and intrinsically deformed states in close proximity \cite{NuDat:2014,Girod:1988fk,Broda:2012zr,Chiara:2012ys,Zhu:2012tg,Recchia:2013kx}.
Traditionally, the onset of deformation is explained by strong quadrupole interactions between nucleons in states with single-particle $\Delta j=2$ and small energy difference. In the reference states for $\nuc{Ni}{64,66}$, the difference between the effective $0f_{5/2}$ and $1p_{1/2}$ single-particle energies \cite{Duguet:2012ys} is merely $200\,\keV$, and therefore sensitive to the balance of NN and 3N tensor and spin-orbit interactions.

\begin{figure}
  \includegraphics[width=0.95\columnwidth]{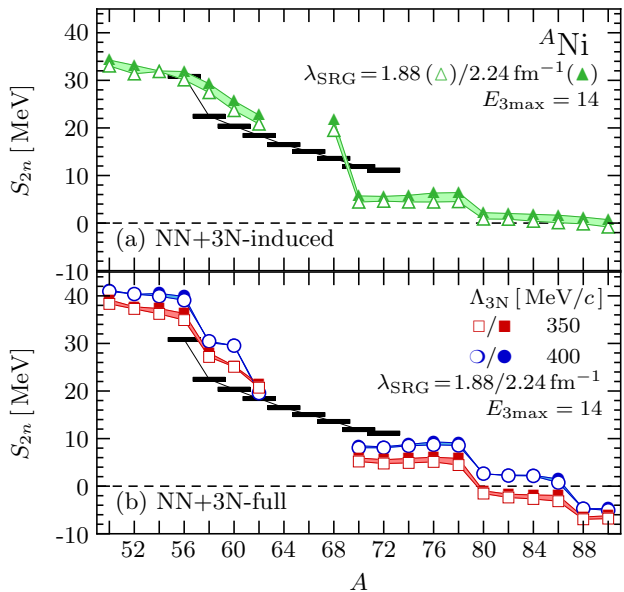}
\caption{\label{fig:Ni_S2N}(Color online) Two-neutron separation energies of the Ni isotopes for the NN+3N-induced (a) and NN+3N-full Hamiltonian with $\Lambda_\text{3N}=350,400\,\MeV/c$ (b), for a range $\lambdaSRG=1.88\fmi$ (open symbols) to $2.24\fmi$ (solid symbols). Experimental values (black bars) are taken from \cite{Wang:2012uq}.}
\end{figure}

The importance of the initial 3N interaction is evident from the $S_{2n}$ shown in Fig.~\ref{fig:Ni_S2N}. Without it (Fig.~\ref{fig:Ni_S2N}(a)), the $N=50$ and especially the $N=28$ shell closures are weak, while the $N=40$ closure is strongly enhanced compared to experiment. Inclusion of the 3N forces (Fig.~\ref{fig:Ni_S2N}(b)) improves the shell closure at $\nuc{Ni}{56}$ significantly, and shifts the $S_{2n}$ of $\nuc{Ni}{70-78}$ closer to experiment. 
Variation of $\Lambda_\text{3N}$ moves the theoretical neuton drip line from $\nuc{Ni}{86}$ to $\nuc{Ni}{78}$, but the $S_{2n}$ are sufficiently small for the situation to change as we improve on the present truncations and include continuum effects. Note also that the experimentally unobserved sub-shell closure in $\nuc{Ni}{60}$ vanishes for $\Lambda_\text{3N}=350\,\MeV/c$. This is a more concrete example of how the internal structure of medium-mass nuclei is affected by variations of $\Lambda_\text{3N}$ than the shift in $S_{2n}$ trends for the Ca isotopes (cf.~Fig.~\ref{fig:Ca_S2N}(b)).

We conclude by discussing the uncertainties of the $S_{2n}$. Using the energies for $\nuc{Ni}{60,62}$, we find a difference of $300-350\,\keV$ between the CCSD and CR-CC(2,3) results, which serves as a measure for the uncertainty due to the many-body truncation. The change of $|\Delta S_{2n}|$ as $\EMax=12\to14\to16$ is similar to Fig.~\ref{fig:Ca_DeltaS2N} for the Ca isotopes. As $\EMax=12\to14$, the change for $\nuc{Ni}{48-86}$ is of the order of $500\,\keV$, beyond that $1-1.5\,\MeV$ for the range $\lambdaSRG=1.88,\ldots,2.24\fmi$. Increasing $\EMax=14\to16$, the change in $S_{2n}$ drops below $250\,\keV$ for $\nuc{Ni}{48-86}$, and to $400-500\keV$ for heavier isotopes.

\paragraph{Conclusions.}
We have studied the even Ca and Ni isotopes with the recently developed MR-IM-SRG, using chiral NN+3N interactions as input. The application of the MR-IM-SRG to the chain of even Ni isotopes marks an important milestone for \emph{ab initio} nuclear structure theory, and shows the viability of such calculations for medium-mass and heavy nuclei. The modest polynomial scaling of the method makes it feasible to reach the tin isotopic chain (and beyond), if sophisticated techniques are implemented to handle or avoid the expensive matrix element storage for 3N interactions \cite{Binder:2014fk}.

The current generation of chiral NN+3N Hamiltonians generally overbind the Ca and Ni isotopes. The fair reproduction of two-neutron separation energies indicates that a good portion of this over-binding amounts to a global shift. We find that an initial 3N interaction is required to reproduce the experimentally confirmed shell closures of $\nuc{Ca}{48,52,54}$. In the Ni isotopes, the creation of an artifical sub-shell closure in $\nuc{Ni}{60}$ and the strong enhancement of the $\nuc{Ni}{68}$ closure indicate that the spin-orbit and tensor components of the chiral 3N interaction might be too strong. Our findings are consistent with earlier studies of medium-mass nuclei based on the same chiral NN+3N Hamiltonians \cite{Roth:2012qf,Binder:2013zr,Hergert:2013mi,Binder:2014fk,Soma:2014eu}, and by extending the range of studied isotopes, we provide further evidence for deficiencies in these Hamiltonians which need to be addressed by the next generation of interactions from chiral EFT.

For neutron-rich Ca isotopes, we predict a very flat trend for the ground-state and two-neutron separation energies, which inhibits a clear identification of the drip line. The interplay of different interaction terms and continuum effects may give rise to interesting physics in this region. Fortunately, these specific Ca isotopes will be investigated in experimental campaigns in the near future. 

\paragraph{Acknowledgments.}
We thank V.~Som\`{a}, T.~Duguet, R. Furnstahl for useful comments and discussions, and V.~Som\`{a} for providing updated GGF results for the Ca isotopes.
Work supported by the National Science Foundation under Grants No.~PHY-1306250, PHY-1068648, and PHY-1404159, and the NUCLEI SciDAC-3 Collaboration under the U.S. Department of Energy Grants No.~DE-SC0008533 and DE-SC0008511. 
Supported by the
Deutsche Forschungsgemeinschaft through contract SFB 634,
by the Helmholtz International Center for FAIR (HIC for
FAIR) within the LOEWE program of the State of Hesse, and
the BMBF through contract 06DA7047I. 
S.~Binder gratefully acknowledges the financial support of the Alexander-von-Humboldt Foundation (Feodor-Lynen scholarship).
Computing resources were provided by the Ohio Supercomputer Center (OSC), the J\"ulich Supercomputing Center, the LOEWE-CSC Frankfurt, and the National Energy Research Scientific Computing Center supported by the Office of Science of the U.S. Department of Energy under Contract No. DE-AC02-05CHH11231.

%

\end{document}